\documentstyle[twocolumn,aps,psfig,amssymb]{revtex}
\input{psfig.sty}
\begin{document}
\title{Harmonics generation in electron-ion collisions in a short laser pulse}
\author{H.~Haberland}
\address{Institut f\"ur Physik, Universit\"at Greifswald, Domstr. 10a,
17487 Greifswald}
\author{M.~Bonitz and D.~Kremp}
\address{Fachbereich Physik, Universit{\"a}t Rostock\\
Universit{\"a}tsplatz 3, D-18051 Rostock, Germany}
\date{\today}
\maketitle
%-------------------------------------------------------------------
\begin{abstract}

Anomalously high generation efficiency of coherent higher field-harmonics in 
collisions between {\em oppositely charged particles} in the field of 
femtosecond lasers is predicted. 
This is based on rigorous numerical solutions of a quantum kinetic equation for 
dense laser plasmas which overcomes limitations of previous investigations.
\end{abstract}
%------------------------------------------------------------------
%-------------------------------------------------------------------
%-------------------------------------------------------------------
%\newpage
Modern high intensity short--pulse lasers \cite{perry-etal.94}
have made it possible to create dense correlated plasmas under laboratory
conditions promising a large variety of applications, e.g. \cite{Tea94,gamaliy}.
Remarkable progress has been made in the field of time--resolved plasma
diagnostics  \cite{theobald-etal.96}, interesting high-field phenomena 
have been observed, including production of multi-keV electrons and the 
generation of higher field harmonics (bremsstrahlung, BS) in neutral gases. 
On the other hand, despite early predictions \cite{silin64}, harmonics generation 
in an {\em ionized} medium due to {\em collisions of oppositely charged particles} 
in the field has not been observed experimentally \cite{gibbon}.  The reason 
is the low electron-ion collision frequency $\nu_{ei}$ in a low density equilibrium 
plasma 
\cite{silin00}. In this Letter we demonstrate a drastic, more than 6 orders of 
magnitude, efficiency increase for dense {\em nonequilibrium plasmas} 
excited by a {\em femtosecond laser pulse} which should make this 
fundamental effect observable experimentally. Moreover, we predict 
coherent harmonics pulses which may be significantly shorter than the 
pump pulse.

A full theoretical description of dense laser plasmas requires the 
selfconsistent treatment of i) hydrodynamic mass and energy transport 
of electrons and heavy particles, ii) collective (mean-field) phenomena and 
iii) elastic and inelastic carrier collisions in the presence of the 
field. In recent years there have been substantial advances in the
theoretical description of the first two effects, 
see e.g. \cite{gamaliy,rozmus90,rozmus96} and \cite{ruhl,pukhov,ruhl99b}, 
respectively.
However, these treatments, usually being based on Vlasov, Fokker-Planck or 
particle in cell 
simulations, either neglect collisions completely or study them in a strongly 
simplified manner. In particular, they neglect the field effect on the collision 
process. This may be justified for very strong fields where 
the electron oscillation velocity $v_0\equiv e E_0/(m_e\Omega)$ is much larger 
than the thermal velocity $v_{th}$, and the collision frequencies $\nu_{ee}$ 
and $\nu_{ei}$ are much below the oscillation frequency $\Omega$ of the field. 
However, for field intensities below $10^{15} {\rm W/cm}^2$) and/or high plasma 
density ($n>10^{20}{\rm cm}^{-3}$) and low temperature, collisions  are the 
dominant factor in field-matter interaction at short times. At a later stage, 
due to hydrodynamic motion of electrons and ions, plasma heating and expansion, 
field effects will govern the plasma behavior, although collisional processes 
such as inverse bremsstrahlung, e.g. \cite{dawson62,silin64,rozmus96} or impact 
ionization/recombination remain important.

{\em When} this transition occurs and {\em what physical processes} 
take place at the {\em early stage} is not known until now - to answer these questions is 
the main goal of this paper. Our analysis reveals that during this initial stage 
the plasma has a strongly anisotropic and non-Maxwellian electron distribution 
function which develops bunches of fast electrons. As a result of electron-electron 
collisions, eventually a monotonically decaying distribution is approached, 
but this thermalization takes about $t_{\rm rel}\sim 50 \dots 100$ fs.
Finally, the most interesting effect is that, under the strong nonequilibrium 
conditions at these short times, the generation of higher field harmonics 
in e-i collisions is strongly enhanced.

The analysis of the initial stage of laser-matter interaction requires 
a quantum kinetic equation which fully includes nonideal plasma effects 
the importance of which has been noted frequently, e.g. \cite{rozmus90,ruhl99b}.
Such an equation has recently been derived \cite{kremp-etal.99pre} and 
has the form

\begin{eqnarray}
&&\left\{\frac{\partial}{\partial t} + e_a {\bf E}(t)\nabla_{\bf k} \right\}
f_a({\bf k}_a,t) = \sum\limits_{b} I_{ab}({\bf k}_a,t),
\label{kin-ea}
\end{eqnarray}
where for the initial period spatial homogeneity may be assumed.
The collision integrals are given by
\begin{eqnarray}
I_{ab}({\bf k}_a,t) = 2
\int \frac{d{\bf k}_b d{\bar {\bf k}}_a d{\bar {\bf k}}_b}{(2\pi\hbar)^6}
|V_{ab}({\bf q})|^2 \delta ({\bf k}_{ab}-{\bar{\bf k}}_{ab})\,
\nonumber\\
\times \int_{t_0}^t d {\bar t} \; \cos\left[
\frac{\epsilon_{ab}-{\bar \epsilon}_{ab}}{\hbar}(t-\bar{t})-
\frac{\bf q}{\hbar}\,{\bf R}_{ab}(t,{\bar t})
\right]
\nonumber\\
\times\left\{{\bar f}_a {\bar f}_b\,[1-f_a] \,[1-f_b] -
f_a f_b \,[1-{\bar f}_a] \, [1-{\bar f}_b]\right\}\big|_{{\bar t}},\;
\label{f_eqfin}
\end{eqnarray}
where $a, b$ label electrons and ions, and we denoted 
$\epsilon_{ab} \equiv \epsilon_{a}+\epsilon_{b}$,
$\epsilon_{a}\equiv p^2_a/2m_a$, ${\bf k}_{ab}\equiv {\bf k}_{a}+{\bf k}_{b}$,
and ${\bf q}\equiv {\bf k}_a - {\bar {\bf k}}_a$.
$V_{ab}$ is the Fourier transform of the screened Coulomb potential
$V_{ab}(q)=4\pi e_a e_b/[q^2+\kappa(t)^2]$, [$\kappa(t)$ is the inverse
{\em nonequilibrium} screening length]. This allows to avoid any cutoff 
(Coulomb logarithm). We underline that the collision integral $I_{ab}$ 
is exact (within the static weak-coupling limit). In particular, it permits
a rigorous investigation of the short-time physics and, for long times
$t \gg 2\pi/\omega_{pl}$ [$\omega_{pl}=(4\pi n e^2/m)^{1/2}$ is the 
plasma frequency] and zero field ($R_{ab}=0$), it goes over to the 
familiar Landau collision integral 
$\int d{\bar t} \cos{...} \rightarrow 
\hbar \pi \delta(\epsilon_{ab}-{\bar \epsilon}_{ab})$. 
Moreover, this integral includes 
Pauli blocking (spin statistics) and it conserves the {\em total energy} 
(kinetic plus potential energy) which is important for dense nonideal plasmas.

Most importantly, the integral $I_{ab}$ includes the {\em influence of 
a strong field on the collision process} of two particles of species $a$ and
$b$  exactly, as well as the effect of a finite collision duration 
$\tau_{\rm coll}=t_2-t_1$
which is crucial for the results shown below.
Indeed, during the time $\tau_{\rm coll}$ each of the colliding particles gains a 
certain momentum ${\bf Q}_a(t_1,t_2)$ in the field, and a pair of colliding
(oppositely charged) particles is displaced by the field a distance
$R_{ab}(t_1,t_2)$. For a harmonic field, ${\bf E}(t)={\bf E}_0\cos \Omega t$, 
\begin{eqnarray}
{\bf Q}_a(t_1,t_2) &=&
\frac{-e_a{\bf E}_0}{\Omega} (\sin \Omega t_1 - \sin\Omega t_2),
\label{qa-def}
\\
{\bf R}_{ab}(t_1,t_2) &=& \left(\frac{e_a}{m_a}-\frac{e_b}{m_b}\right)
\frac{{\bf E}_0}{\Omega^2}
\nonumber\\
&\times&
\left\{\cos \Omega t_1-\cos \Omega t_2-
\Omega\,(t_2-t_1)\sin\Omega t_1\right\}, 
\label{rab-def}
\end{eqnarray}
and these effects are growing with increasing field strength.
The momentum shift enters the arguments of the distribution functions in 
the collision integral (\ref{f_eqfin}), 
(time-dependent intra-collisional field effect):
$f_a \equiv f[{\bf k}_a+{\bf Q}_a(t,{\bar t}),{\bar t}]$;
$\:{\bar f}_a \equiv f[{\bar {\bf k}}_a+{\bf Q}_a(t,{\bar t}),{\bar t}]$,  
whereas ${\bf R}_{ab}$ modifies the energy balance of the two-particle
collisions in a monochromatic field. Thus the collision integral (\ref{f_eqfin}) 
selfconsistently
includes nonlinear field effects, such as harmonics
generation, emission (absorption) of laser photons during the two-particle
scattering [(inverse) BS] and generalizes previous treatments 
of these phenomena, e.g. \cite{silin64,rozmus96}, to the case of arbitrary 
short times, nonequilibrium distributions and modifications by dense 
plasma effects.

We expect the most interesting physics to occur in situations where the effects 
of collisions and of the field are of the same order: (i) when the field 
frequency $\Omega$ is in the range of the collision frequency $\nu_{ei}$ and 
the plasma frequency $\omega_{pl}$ and, (ii) when the 
field induced particle velocities $v_0$ are of the same order as the thermal 
velocity $v_{th}$. Thus, considering as an example a dense fully ionized
hydrogen plasma in a  monochromatic optical field of intermediate wavelength
$\lambda=500$nm  ($\Omega=3.77 {\rm fs}^{-1}$), the following 
parameter range is of interest for the simulations:
 densities $10^{20}{\rm cm}^{-3}\le n \le 10^{24}{\rm cm}^{-3}$,
corresponding to a plasma frequency range 
$0.56~{\rm fs}^{-1} \le \omega_{pl}\le 56~{\rm fs}^{-1}$, and an initial plasma 
temperature of $20,000K$ which may increase up to almost $10^6$K within the first 
$50$fs of the relaxation. Correspondingly, for this case field strengths of 
interest are $10^7$V/cm $\ge E_0 \ge 10^9$V/cm. For stronger fields 
($I >10^{15}$W/cm$^{-2}$) collisions will already be of minor importance.

We solve equations (\ref{kin-ea}-\ref{rab-def}) by direct numerical integration 
starting from a pre-excited electron-ion plasma in equilibrium, 
and the relaxation is computed over about 50 field cycles. We underline that 
we solve equations (\ref{kin-ea}-\ref{rab-def}) without any simplifying 
assumptions. In particular, momentum anisotropy  of the 
electron distribution function as well as time and momentum retardation
are fully included. The ion distribution was found to remain a Maxwellian 
during the considered time interval which allowed us to simplify the integral 
$I_{ei}$.
To properly account for the time-dependence of the screened Coulomb
potential $V(q)$, the inverse screening length $\kappa(t)$ is computed 
selfconsistently from the current nonequilibrium distribution functions. 
Solving the kinetic equation yields the time and momentum-dependent
distribution function $f_e({\bf k},t)$ which allows us to compute all
transport properties, including the mean electron kinetic energy density
$\langle E^{e}_{\rm kin} \rangle (t) =
\frac{m_e}{2} \langle v_e^2 \rangle$ and
the electrical current density 
${\bf j}(t) =  \sum_a e_a \langle {\bf v}_a \rangle$, 
[$f$ is normalized to the density, and $\langle \dots \rangle$ denotes 
averaging over $f(t)$].
\vspace{1.3cm} 

\begin{figure}
\hspace{0.3cm}
\psfig{figure=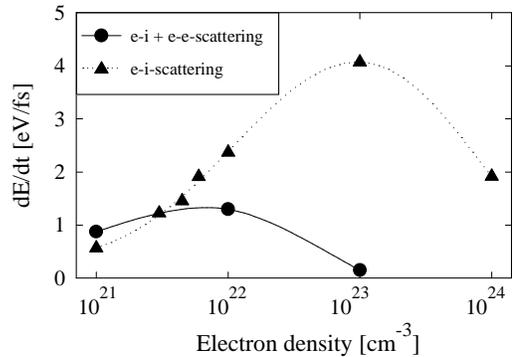,width=7cm,angle=0}
\vspace{0.5cm} 
\caption{Average electron kinetic energy increase
versus density, with and without e-e scattering included. 
Field amplitude and wavelength are $E_0=3 \times 10^8 {\rm V/cm}$, 
$\lambda=500{\rm nm}$, the initial plasma temperature is 
$T_0=20,000{\rm K}$.}
\label{fig1}
%\vspace{0.5cm}
\end{figure}

Let us now come to the results. First recall that, in the collisionless 
case ($I_{ab}=0$), the electron distribution would oscillate as a whole 
according to $\sin \Omega t$ without changing shape. This yields no net 
current and no energy increase (the energy oscillates with $2\Omega$ around
the ponderomotive energy $U_{\rm pond}=m v_0^2/4$).
In contrast, in the presence of collisions, we observe a strong increase of 
the mean electron kinetic energy. After an initial transient of about
$5~{\rm fs}$, $\langle E^e_{\rm kin} \rangle $ grows almost linearly, thereby
oscillating with $2\Omega$. 
This energy increase is due to {\em field-induced 
electron acceleration during electron-ion collisions}, cf. the term 
${\bf q R}_{ei}$ in the integral $I_{ei},$ Eq.~(\ref{f_eqfin}). No net 
acceleration occurs in electron-electron collisions (${\bf R}_{ee}=0$).
The energy growth rate depends on the ratio of 
field strength and density. This is shown in Fig.~\ref{fig1} where the average 
slope of the energy curves is plotted for different densities. 
Interestingly, the most efficient plasma heating occurs around 
$10^{22}{\rm cm}^{-3}$ which is near the critical density 
$n_{cr}=4.5 \cdot 10^{21} {\rm cm}^{-3}$
where $\omega_{pl}=\Omega$. The maximum value of the power is about
$1.5$eV/fs which roughly corresponds to the absorption of one
photon per electron and laser cycle. 
Notice also that, although e-e scattering does not contribute to the heating 
directly, this process may not be neglected, as this would
lead to an overestimate of the power
by a factor of three and a shift of the maximum to higher densities by one
order of magnitude, cf. the triangles in Fig.~\ref{fig1}.

\begin{figure}
\psfig{figure=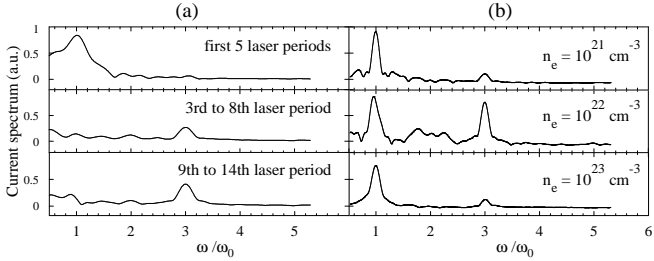,width=8.7cm,angle=0}
\vspace{0.5cm} 
\caption{Electrical current spectrum in a time-dependent 
electric field. The higher harmonics are the sole result of electron-ion 
collisions. (a) - time evolution of the spectrum for $n=10^{22}{\rm cm}^{-3}$, 
from averaging over 5 laser periods (see text in fig.).
(b) - spectrum (from averaging over the full calculation) for 
various densities. Other parameters are the same as in Fig.~\ref{fig1}.}
\label{fig2}
\end{figure}

Next, we consider the electrical current density.
In contrast to the collisionless case, where 
${\bf j}_0(t)=e n {\bf v}_0 \sin \Omega t$, 
collisions between electrons and ions may give rise to higher field harmonics. 
Indeed, electrons oscillating with the field are disturbed by the Coulomb field 
of ions which yields a change of the electron velocity component in field direction 
($v_0 \gg v_{th}$)
\begin{eqnarray}
\Delta v_z(t) &\approx& \frac{a \,\sqrt{2}}{\Omega\,(1+b^2)}
\frac{\sin {\Omega t}}{\sqrt{1+2 b^2 + \cos{2\Omega t}}},
\label{dv}
\\
\mbox{with} \quad a &=& \frac{e_e e_i}{\mu r_0^2},\;
b=\sqrt{\frac{r_{\bot}}{r_0}}, \; 
r_0=\left(\frac{e_e}{m_e}-\frac{e_i}{m_i}\right)\frac{E_0}{\Omega^2},
\nonumber
\end{eqnarray}
where $\mu^{-1}=m_e^{-1}+m_i^{-1}$, and $r_{\bot}$ is the minimal 
electron-ion distance. One readily checks that in the spectrum of 
(\ref{dv}) appear 
all harmonics of the field. Formula (\ref{dv}) indicates that the 
spectrum will be particularly broad for close encounters (small $r_{\bot}$, 
strong scattering). Further, the efficiency of harmonics generation
decreases with growing field strength. On the other hand, for the case of 
a weak field [not covered by Eq. (\ref{dv})], 
the harmonics yield increases with $E_0$ (with the number of photons), 
resulting in a maximum of the efficiency around $v_0 = v_{th}$ 
\cite{silin00,paul}.
% and increases with density 
%(growing frequency $\nu_{ei}$ of scattering events). 
Quantitative estimates for the conversion efficiency $\eta_N$ (intensity 
of the $N$-th harmonics normalized 
to the intensity of the fundamental oscillation) 
can be given for the case of Maxwellian electrons.
Then only odd harmonics exist and 
$\eta^{EQ}_{N} \sim (\nu_{ei}/\Omega)^2 /N^{\alpha_N}$, where 
$\alpha_N=\alpha_N(v_0/v_{th}) > 0$ \cite{silin00}. Direct evaluation of 
the collision integral $I_{ei}$, Eq. (\ref{f_eqfin}), for a dense 
equilibrium plasma revealed that the highest value in equilibrium is 
$\eta^{EQ}_3 \sim 10^{-7}$ \cite{paul}.

\begin{figure}
\psfig{figure=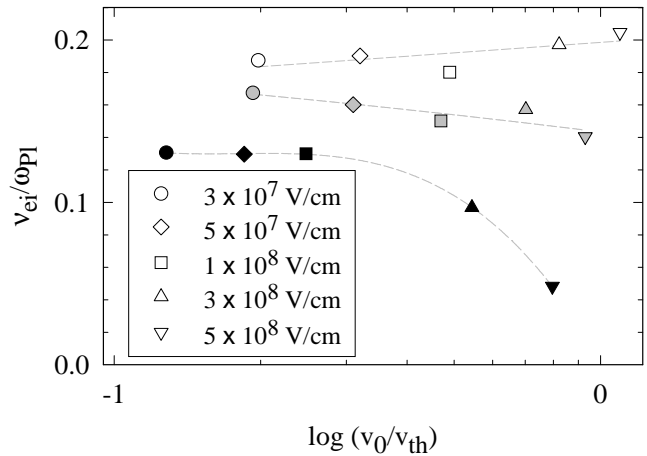,width=8.5cm,angle=0}
\vspace{0.5cm} 
\caption{Nonequilibrium collision frequency, Eq. (\ref{nu_ei}), 
at the initial moment, after 9fs and after 25 fs (white, grey and 
black symbols, respectively). 
$n=10^{22}{\rm cm}^{-3}$, $v_{th}\equiv \sqrt{\langle v_e^2\rangle}$. 
Due to collisional energy absorption, cf. Fig.~1, during the relaxation,
symbols  move to the left. Symbols with same shape refer to the same field
strength, lines are guides to the eye.}
\label{fig3}
\end{figure}

However, as we show now, the efficiency may be increased drastically under 
nonequilibrium conditions of femtosecond laser pulse excitation. 
Fig.~\ref{fig2} shows the results for the current spectrum obtained from the 
numerical solution of the quantum kinetic equation (\ref{kin-ea},\ref{f_eqfin}) 
for $E_0=3\times 10^8 {\rm V/cm}$ and an initial temperature $T=20,000$K.
Fig.~\ref{fig2}.a shows the result for a density of $n=10^{22} {\rm cm}^{-3}$ at 
different times (obtained by Fourier transforming ${\bf j}(t)$ over time 
intervals of 5 laser periods). One clearly sees the formation of a strong third 
harmonics which is comparable to and may even exceed the fundamental component, 
i.e. $\eta_3 \sim 1$. 
At later times, this peak vanishes again.
We mention that higher odd harmonics are found for higher field strength also,
besides, there is a clear signature of the second harmonics, cf. Fig.~\ref{fig2}. 
The temporal behavior of the harmonics emission can be understood from the 
time evolution of the nonequilibrium cycle-averaged collision frequency 
\begin{eqnarray}
\nu_{ei}(E_0, \Omega; n, t)=4\pi\frac{\Omega^2}{\omega_{pl}^2}\,
\frac{\overline{{\bf j}(t){\bf E}(t)}}{\overline{{\bf E}^2(t)}},
\label{nu_ei}
\end{eqnarray}
which is shown in Fig.~\ref{fig3}.
Due to the laser heating of the plasma, cf. Fig.~\ref{fig1}, and relaxation 
towards a Maxwellian, the collision 
frequency $\nu_{ei}$ decreases in time, limiting the efficiency 
$\eta_N(t)$ and, thus, the duration of the harmonics pulse. 

\begin{figure}
\psfig{figure=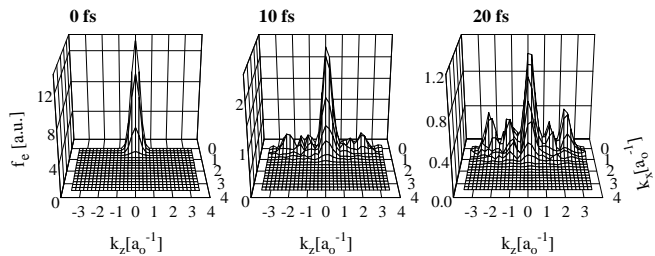,width=8.7cm,angle=0}
\vspace{0.5cm} 
\caption{Evolution of the electron distribution function in
momentum space (the field is in $k_z$ direction).
Figure shows snapshots at
$t=0$ (left), after 6 laser periods (middle)
and after 12 laser periods (right), when the main peak of $f_e$ is at the origin
(the distribution as a whole oscillates with the field 
in $k_z$-direction and is isotropic in the $k_x$-$k_y$-plane).
Note the different vertical scales.}
\label{fig4}
\end{figure}

Finally, to understand the 
reason for the anomalously high efficiency of harmonics generation, we consider 
in Fig.~\ref{fig4} the temporal evolution of the electron distribution. 
The three figure parts show snapshots of $f_e(k_x,k_z)$ after $0, 6$ and $12$ 
complete field cycles.
Clearly, the distribution is becoming  anisotropic 
already during the first few laser periods. It strongly deviates from a 
Maxwellian, developing side peaks along the field direction, the distance 
of which from the origin is proportional to the field strength. This 
peculiar nonequilibrium behavior is caused by electron-ion collisions 
in the presence of the field. It is observed for all considered densities and 
field strengths (see above), but is strongest around $\Omega=\omega_{pl}$ 
and $v_0=v_{th}$. Under these conditions, an electron spends the maximum possible
time in the field of an ion [extending over approximately $r_D=\kappa^{-1}$] 
per laser cycle, 
$\Omega\tau_{\rm coll} \sim \Omega r_D/v_{th} \sim \Omega/{\omega_{pl}}\sim 1$.
These are the optimal conditions for collisional e-i momentum exchange in the
field, as for electron acceleration (inverse BS), cf. Fig.~\ref{fig1}, and 
harmonics emission (BS), Fig.~\ref{fig2}.b.

Our calculations show that, as a result of
e-e collisions, the side peaks of $f_e$ start to smear out after 
$30 \dots 50$fs, and after $t_{\rm rel} \sim 50 \dots 100$fs (depending on the density) 
a monotonically decaying distribution with a temperature of the order of several 
$10^6 K$ is reached. By this time the plasma has become essentially collisionless: 
$\nu_{ei}$ has decayed to its equilibrium value \cite{bornath-etal.00lpb} 
and $\nu_{ee} \approx 0$, making it necessary to include collective and 
hydrodynamice (spatial heat flow, expansion etc.) effects, 
as mentioned in the Introduction, while for the treatment of collisions simpler
models may be used.

In summary, we have presented a selfconsistent numerical investigation of the 
early relaxation stage, $t\le 50$fs, of a pre-ionized dense plasma in a laser 
field with $E_0=10^7\dots 10^9 {\rm V/cm}$.. We have found the formation of a 
strongly nonequilibrium electron distribution giving rise to 
coherent bremsstrahlung emission of ultrashort third harmonics pulses.
An anomalously high conversion efficiency (more than 6 orders of magnitude 
compared to plasmas in equilibrium) is predicted, making this fundamental process 
observable in experiments. The most favorable conditions are expected for a 
pre-ionized weakly undercritical gas plasma with high charge state
($\nu_{ei}\sim Z$) excited by a linearly polarized laser  pulse with a duration
$\tau_p$ shorter than $50$fs where the third harmonics  should be observable in
transmission. To further improve the quantitative predictions, the plasma
generation  process has to be included in the simulation. For example,
ionization  from excited atomic levels should enhance the  harmonics yield 
\cite{silin00}, accelerate the plasma heating and thus further
 reduce the duration of the higher harmonics pulse.
Finally, we  mention that our results are not limited to laser plasmas, but
are equally  important for electron-hole plasmas subject to intense THz
radiation.

This work is supported by the Deutsche Forschungsgemeinschaft
(Schwerpunkt ``Laserfelder''), the European Commission through the
TMR Network SILASI and the NIC J\"ulich.
We acknowledge stimulating discussions with M.~Schlanges and V.P.~Silin.

%---------------------------------------------------------------------------------


\begin{references}

\bibitem{perry-etal.94}
for a recent overview, see e.g. M.D.~Perry, and G.~Mourou,
Science {\bf 264} (1994) 917

\bibitem{Tea94} M.~Tabak et al.,
Phys. Plasmas {\bf 1} (1994) 1626

\bibitem{gamaliy} Gamaliy, Laser \& Particle Beams {\bf 12}, 185 (1994)

\bibitem{theobald-etal.96}
W.~Theobald, R.~H\"assner, C.~W\"ulker, and R.~Sauerbrey,
Phys. Rev. Lett. {\bf 77}, 298 (1996) 

\bibitem{silin64} V.P.~Silin, Sov. Phys. JETP {\bf 20}, 1510 (1965)

\bibitem{gibbon} The BS mechanism is fundamentally 
different from the 
collisionless harmonics generation in relativistic plasmas, 
see e.g. P.~Gibbon, J. Quantum Electronics (1996) and refs. therein.

\bibitem{silin00} See V.P.~Silin, Sov. Phys. JETP {\bf 90}, 805 (2000);
Kvantovaya Elektronika (russ.) {\bf 26}, 11 (1999).

\bibitem{rozmus90} W.~Rozmus, and V.T.~Tikhonchuk,
Phys. Rev. A {\bf 42}, 7401 (1990)

\bibitem{rozmus96} For a recent overview and further references, see
D.~Vick, C.E.~Capjack, V.~Tikhonchuk, and W.~Rozmus,
Comments Plasma Phys. Controlled Fusion {\bf 17}, 87 (1996)

\bibitem{ruhl} H.~Ruhl et al., Phys. Rev. Lett. {\bf 82}, 743 (1999)

\bibitem{pukhov} A.~Pukhov, and J.~Meyer-ter-Vehn,
Phys. Rev. Lett. {\bf 76}, 3975 (1996)

\bibitem{ruhl99b} H.~Ruhl et al., Phys. Rev. Lett. {\bf 82}, 2095 (1999)

\bibitem{dawson62} J.~Dawson, and C.~Oberman,
Phys. Fluids {\bf 5}, 517 (1962)

\bibitem{kremp-etal.99pre} D.~Kremp, Th.~Bornath, M.~Bonitz, and
M.~Schlanges, Phys. Rev. E {\bf 60}, 4725 (1999).

\bibitem{paul} P.~Hilse, to be published

\bibitem{bornath-etal.00lpb} Th.~Bornath, M.~Schlanges, P.~Hilse,
D.~Kremp, and M.~Bonitz, Laser \& Particle Beams (2000)

\end{references}
\end{document}